\documentclass[amsmath,amssymb,prl,twocolumn,superscriptaddress]{revtex4}

\usepackage[utf8]{inputenc}
\usepackage{amsmath}
\usepackage{graphicx}
\usepackage{chemformula}
\usepackage{hyperref}
\usepackage{enumerate}
\usepackage{xcolor}
\newcommand{\ciceco}{CICECO - Instituto de Materiais de Aveiro, Department of Chemistry, 3810-193 Aveiro, Portugal}

\begin{document}
\title{Organic altermagnets based in two-dimensional nanographene frameworks}

\author{Ricardo Ortiz}\thanks{Corresponding author: ricardo.ortiz.cano@ua.pt}
\affiliation{\ciceco}
\author{Karol Strutyński}
\affiliation{\ciceco}
\author {Manuel Melle-Franco}
\affiliation{\ciceco}

\begin{abstract}
Altermagnetism stands as a third type of collinear magnetic order, whose band structure combines a net zero magnetization with a non-relativistic spin-splitting caused by a broken time reversal symmetry. So far, the strategy to design platforms displaying altermagnetism has relied mostly on inorganic crystals with $d$-metals as spin centers, where a representative example is the two-dimensional square lattice with antiparallel $D_{2h}$ magnetic blocks related by a $\pi/2$ rotation. Despite the fact that there is no strong requirement for the magnetic atoms to be metals, the construction of an altermagnetic framework with light elements like carbon is challenging due to symmetric constrictions. We show how it is possible to overcome this by including non-alternant rings in $\pi$-conjugated nanographenes. More specifically, dibenzo[ef,kl]heptalene, an $S = 1$ $\pi$-conjugated hydrocarbon consisting of a graph of two fused heptagons and hexagons, represents a suitable building block for an altermagnetic 2D crystal. In this work, we confirm this hypothesis with DFT calculations of the spin polarized band structure, presenting a spin compensated ground state with broken time reversal symmetry, and a d-wave symmetry of the first valence and conduction bands. Consistent results are obtained for covalent organic frameworks based on dibenzo[ef,kl]heptalene units connected by linkers, paving the way for the realization of organic altermagnetic materials. 
\end{abstract}
\maketitle

\section{Introduction}

The production and manipulation of spin currents is the cornerstone of prototypical technology based on spintronics\cite{HIROHATA2020166711}. For this matter, in the last decades there has been a growing interest in the fine control of spin polarization,
leading to a whole ecosystem of methodologies\cite{PhysRevB.63.054416,mathon2001theory,konig2007quantum,loss1992persistent,tashiro2015spin}. In this sense, altermagnets have emerged as a promising material for spintronics, since their band structure with spin-split Fermi surfaces makes them act as spin-splitters, producing a transverse spin current as response to a bias applied diagonally to the anisotropy axes\cite{gonzalez2021efficient,vsmejkal2022emerging,fender2024altermagnetism}. 

In altermagnetism (AM), adjacent magnetic moments with equal magnitude are antiparallel to each other, but unlike antiferromagnets, the AM phase is not invariant under $\bold{t}_{\frac{1}{2}}{\cal T}$ or ${\cal P}{\cal T}$ symmetry operations\cite{vsmejkal2020crystal,hayami2019momentum,yuan2020giant,mazin2021prediction}, where ${\cal T}$ is time reversal symmetry, $\bold{t_{\frac{1}{2}}}$ a half unit translation, and ${\cal P}$ inversion symmetry. In addition to these, the two spin sublattices relate by a real space crystal rotation (${\cal A}$), which triggers the lifting of the Kramers degeneracy in some regions of the reciprocal space\cite{vsmejkal2020crystal} ($E_{\bold{k},\sigma}\neq E_{\bold{k},\bar{\sigma}}$). As a consequence, AM presents a combination of features from ferro- and antiferromagnets: a band structure with a non-relativistic spin splitting (NRSS), and a compensated spin polarization.

In two dimensions, spin-split bands can also be found in relativistic materials with a non-negligible spin-orbit coupling (SOC) caused by the Bychkov-Rashba effect\cite{bychkov1984properties,bihlmayer2022rashba}. Such spin-momentum locking produces the Edelstein effect in the presence of a charge current\cite{edelstein1990spin}, generating a spin accumulation with no need of an external magnetic field. Another example are materials with a strong SOC like \ch{Pt}, where transverse spin polarized currents can be produced by the spin Hall effect\cite{sinova2015spin}. However, since SOC scales with the $Z$ atomic number\cite{shanavas2014theoretical}, the optimization of these effects happens at the expense of relying on heavy elements.
The NRSS in AM, as it is confirmed by calculations\cite{vsmejkal2020crystal} and minimal models\cite{das2024realizing,roig2024minimal}, does not require SOC at all, since it is the lattice geometry that breaks ${\cal T}$ in the magnetic phase. 

The typical design of altermagnetic materials consists of the combination of $d$-metals as spin centers and non-magnetic atoms as connections that introduce the rotation symmetry\cite{vsmejkal2022emerging,fender2024altermagnetism}. The rutile lattice, i.e. RuO$_2$, is the archetypical altermagnet, hosting a non-negligible NRSS in the spin polarized band structure\cite{vsmejkal2020crystal}. Currently, its ground state is under debate by experiments that claim the absence of magnetism\cite{hiraishi2024nonmagnetic,kiefer2025crystal} or NRSS\cite{liu2024absence}, in contrast with experiments that report anomalous Hall effect\cite{feng2022anomalous}, magnetic order\cite{berlijn2017itinerant,zhu2019anomalous}, spin-transfer torque\cite{bose2022tilted} and a broken ${\cal T}$\cite{fedchenko2024observation}.
Nonetheless, additional examples of altermagnetic candidates, both 2D and 3D, can be found in the literature, like MnTe\cite{krempasky2024altermagnetic}, RuF$_4$\cite{milivojevic2024interplay}, perovskites\cite{bernardini2025ruddlesden} or twisted magnetic van der Waals bilayers\cite{liu2024twisted}, among others\cite{bai2024altermagnetism,vina2025building}, most of them containing metal atoms as the irremediable source of magnetism. 

\begin{figure*}
 \centering
    \includegraphics[width=0.9\textwidth]{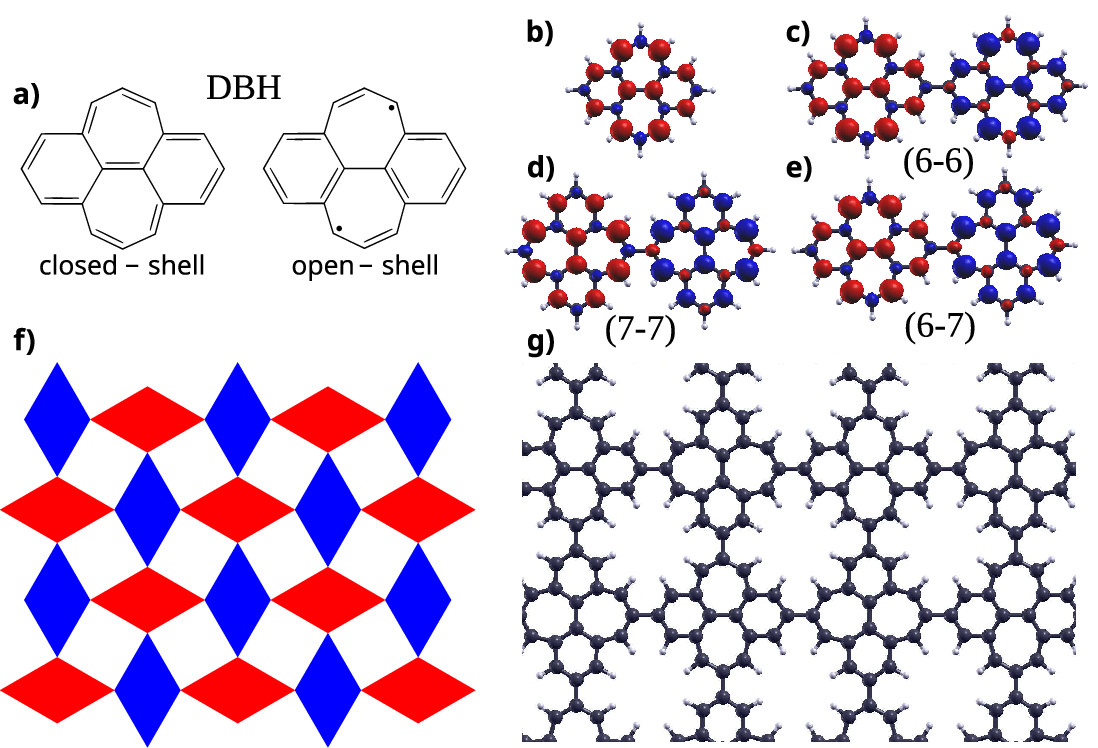}
\caption{a) Closed-shell and open-shell chemical structures of DBH. b) Spin density of the $S=1$ planar DBH monomer. c)-e) Spin density of the three open-shell $S=0$ planar DBH dimers. f) Schetch of an altermagnetic lattice. g) Planar DBH two-dimensional crystal. The calculations of the spin densities are done with DFT with the BP86 density functional and 6-31G** basis. Red and blue stand for the sign.}
\label{fig1}
\end{figure*}

Recently, there has been a bloom in the production of open-shell nanographenes with on-surface synthesis techniques allowing the exploration of magnetism emerging from $\pi$-orbitals\cite{Carbonbasednanostructures}. Different to $d$-magnetism, magnetic nanographenes present a spin density that is delocalized over several carbon atoms, while at the same time it is localized at certain regions of the molecule like, for instance, zigzag edges\cite{JFR07}. This has its origin in the presence of localized in-gap molecular orbitals that host unpaired electrons, making the molecule prone to magnetism by a Stoner instability\cite{Yazyev_2010}. 
In the particular case of alternant nanographenes, the Lieb's theorem\cite{Lieb89} accurately predicts the spin quantum number of the ground state in terms of the imbalance between both sublattices ($S=|N_A-N_B|/2$).     
  
Such a fine tuning of the magnetic properties by the molecular structure, in combination with a low SOC, turns open-shell nanographenes into versatile platforms to study collinear magnetism and spin dynamics. In this sense, different experimental groups have been able to obtain nanographenes displaying ferro- or antiferromagnetism, like triangulenes\cite{pavlivcek2017synthesis,mishra2019synthesis,su2019atomically,mishra2021synthesis} or the Clar's goblet\cite{mishra19b}, whose magnetic fingerprints were probed by measuring a Kondo peak\cite{NachoNat,jacob2021renormalization} or inelastic steps\cite{rhombenes,ORTIZ2020100595} with STM spectroscopy. More complex architectures, where two or more nanographenes are put together\cite{cheng2022surface,paschke2025route}, have also been fabricated, opening the door to non-trivial physics like spin fractionalization in $S=1$ triangulene chains\cite{mishra2021observation}. 

While 2D $\pi$-magnetism is, at this moment, becoming more prominent theoretically, unquenched local moments are widely expected to survive in 2D open-shell nanographene crystals\cite{2Dtriang,henriques2024beyond}.
In this work, we design a 2D framework with $S=1$ nanographenes as building blocks, where the inclusion of non-alternant rings permits the construction of a 2D lattice symmetrically akin to rutile with ${\cal A} = C_{4}$, avoiding the geometric restrictions of alternant nanographenes, and displaying an NRSS.   

\section{The DBH molecule}

Regardless of the chemical structure, alternant nanographenes containing exclusively hexagonal rings cannot be covalently bonded by $sp^2$ carbon atoms if one of them is rotated $90^\circ$ with respect the other. Consequently, a 2D square lattice made of nanographenes connected by a $\pi/2$ rotation is not stable for this kind of molecules. For instance, pyrene is a planar molecule with $D_{2h}$ symmetry, but since it only hosts hexagonal rings, then it is not possible to link two $\pi/2$ rotated pyrenes unless the $sp^2$ hybridization of a carbon atom changes to $sp^3$. However, this can be overcome by non-alternant nanographenes with odd-numbered rings, which are also known for displaying magnetism\cite{liu2025steering,biswas2022defect,zhou2025triplet,ortiz2023magnetic,mishra2024bistability}. In the case of pyrene, substituting the two central hexagons by two heptagons produces a molecule known as dibenzo[ef,kl]heptalene (DBH), which can be linked to neighbouring units by only one bond in each of the four directions of a square lattice.  

 \begin{table}[htp]
 \begin{center}
    \begin{tabular}{ c | c | c | c | c | c | c |}
    \cline{2-7}
   & \multicolumn{6}{c|}{$\Delta E$ (meV)}  \\
    \cline{2-7}
   & \multicolumn{3}{c|}{BP86/6-31G**} & \multicolumn{3}{c|}{PBE0/def2-SVP} \\
    \hline
    \multicolumn{1}{|c|}{Molecule} & $S=0$ & $S=1$ &  $S=2$ & $S=0$ & $S=1$ &  $S=2$ \\ \hline
    \multicolumn{1}{|c|}{DBH} & 233 & 0 & - & 225 & 0 & - \\ 
    \multicolumn{1}{|c|}{DBH dimer (6-6)} & 0 & 9  & 20 & 0 & 9 & 21 \\ 
    \multicolumn{1}{|c|}{DBH dimer (7-7)} & 0 & 25 & 102 & 0 & 27 & 114 \\ 
    \multicolumn{1}{|c|}{DBH dimer (6-7)} & 0 & 12 & 35 & 0 & 16 & 47 \\
    \hline

    \end{tabular}
    \label{table1}
\end{center}
\caption{Lowest energy many-body states referred to the ground state ($\Delta E$) for the planar molecules of Figure \ref{fig1}b-e, calculated with the CASSCF-NEVPT2 methodology. The geometries used were relaxed with DFT for the magnetic solutions shown in Figure \ref{fig1}, using BP86/6-31G** and PBE0/def2-SVP (left and right columns, respectively).}\label{table1}
 \end{table} 


DBH is a non-alternant $\pi$-conjugated nanographene
that consists of two fused heptagons with two hexagons. In
vacuum, DFT calculations predict an $S = 1$ non-planar ground-state. The molecular structure is in line with previous experimental observations on solution closed-shell derivatives\cite{pascal1992synthesis,rashidi2002stereochemistry}.
It is, however, for the sake of simplicity, rather convenient to study first the emergence of AM on 2D flat crystals.

Planar DBH (hereafter just DBH) hosts two Clar's sextets and two unpaired electrons (Fig.\ref{fig1}a). DFT calculations suggest that the ground state presents a ferromagnetic spin density (Fig.\ref{fig1}b), while sophisticated CASSCF-NEVPT2 calculations confirm a spin triplet as the state with lowest energy (Table \ref{table1}). Such a high-spin ground state can be easily rationalized if we consider a DBH molecule without the bond that joins both hexagons, recovering the sublattice symmetry with $|N_A-N_B|=2$, and an $S=1$ ground state according to Lieb's theorem.

 \begin{figure}[h!]
 \centering
    \includegraphics[width=0.5\textwidth]{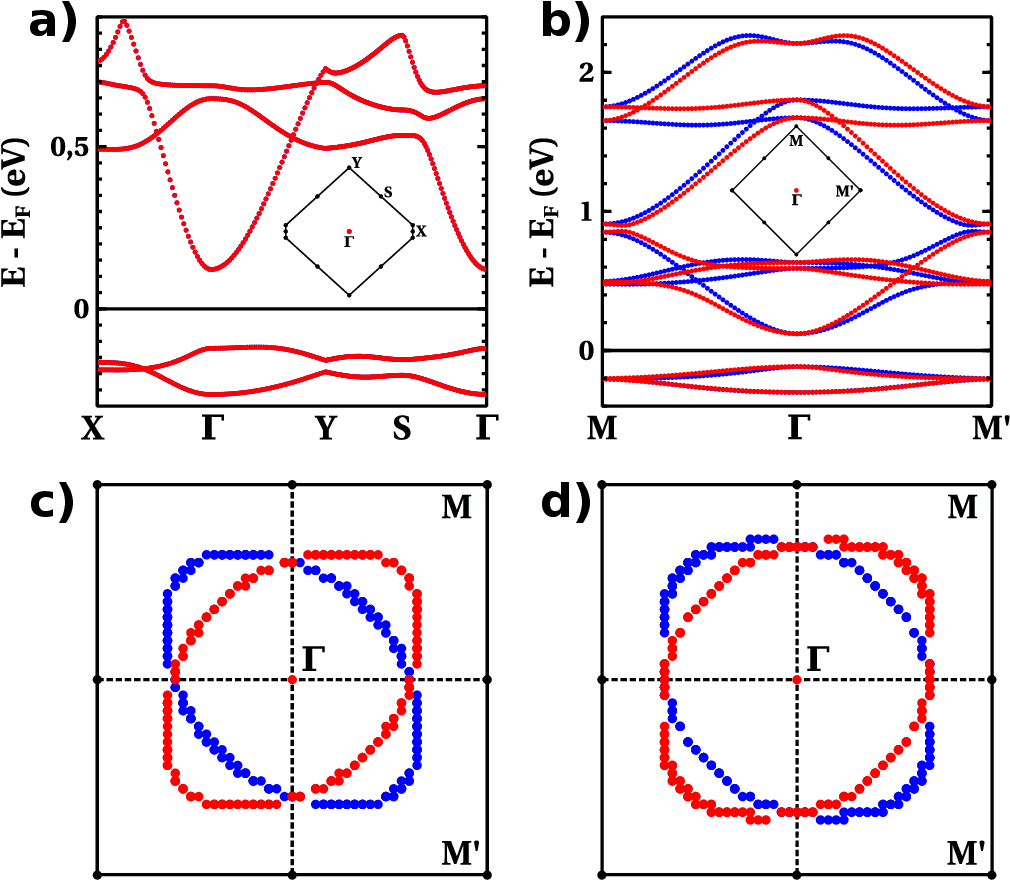}
\caption{Calculated DFT band structure of a) ${\cal A} = C_2$ antiferromagnetic and b) ${\cal A} = C_4$ altermagnetic planar DBH crystals, referred to the Fermi energy ($E_F=-2.855$eV and $E_F=-2.83$eV, respectively). Insets are the first Brillouin zones with labels for the high symmetry points. Constant energy contours of the band structure in b) with energy ranges c) $-3.000\pm0.002eV$ and d) $-2.40\pm0.01eV$ .  Red and blue colors stand for the spin sign. The red dot in the center of the Brillouin zones and contours indicates the $\Gamma$ point.}
\label{fig2}
\end{figure}

The $D_{2h}$ symmetry of DBH permits the formation of dimers just by linking the vertices of either two hexagons, two heptagons or a hexagon with a heptagon. By inspection of the SONOs wave functions from different units, coupling happens mainly at third nearest-neighbours, which produces a kinetic superexchange that is antiferromagnetic\cite{jacob2022theory}. In Figure \ref{fig1}c-e we show that the spin density survives in DBH dimers with antiparallel alignment, while the many-body ground state is an open-shell $S=0$, followed by $S=1$ and $S=2$ excited states (Table\ref{table1}). If we consider both DBH as $S=1$ spins, then we have the following Heisenberg Hamiltonian:

\begin{equation}
{\cal H} = J\bold{S}_1\cdot\bold{S}_2,
\label{eq1}
\end{equation}

where $J>0$ is an antiferromagnetic exchange, and $\bold{S}_{1,2}$ are $S=1$ Pauli matrices vectors. The spin Hamiltonian from Equation \ref{eq1} has an $S=0$ ground state, also followed by $S=1$ and $S=2$ excited states\cite{mishra2020}, separated by $E_{S=1} - E_{S=0} = J$ and $E_{S=2}-E_{S=1} = 2J$, in qualitative agreement with CASSCF-NEVPT2 calculations of DBH dimers (Table \ref{table1}). In consequence, a periodic bidimensional framework made of DBH with ${\cal A} = C_{4}$ is expected to behave as an altermagnetic crystal (Fig.\ref{fig1}f,g).

\section{Altermagnetism in the planar DBH 2D framework}

  \begin{figure}[h!]
 \centering
    \includegraphics[width=0.5\textwidth]{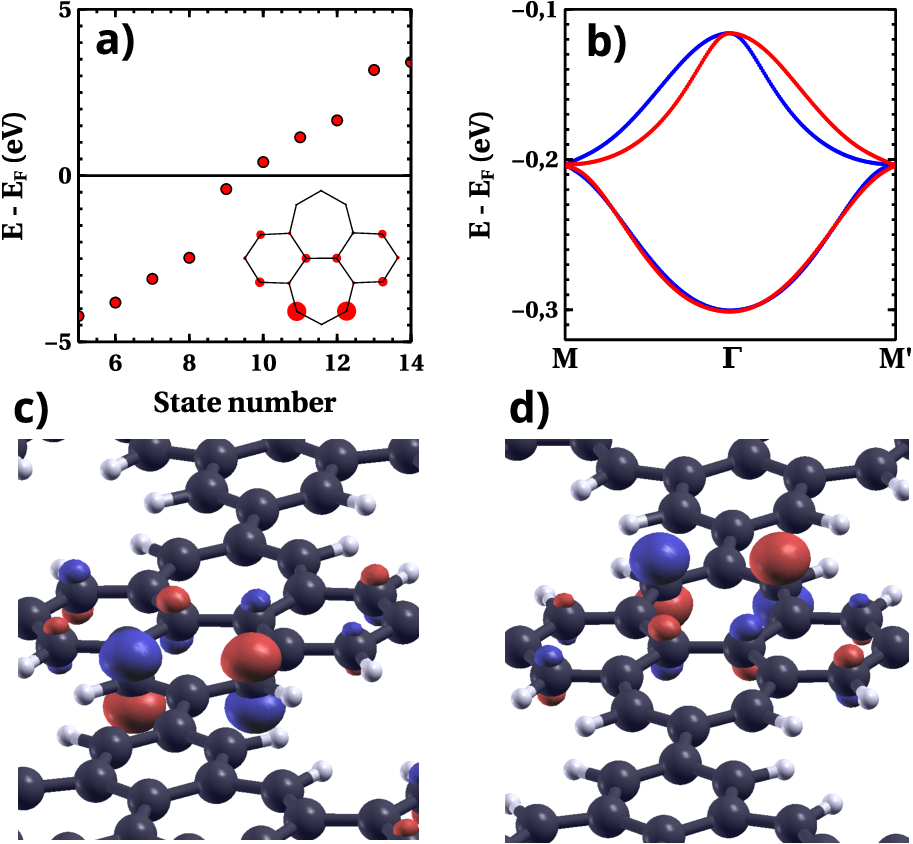}
\caption{a) Tight-binding eigenvalue spectrum of planar DBH with $t=-2.7$eV and $E_F = 0.4065$eV. The inset corresponds to the wave function  of an unpaired electron $|\varphi_{z}(i)|^2$. b) Four highest occupied Wannier bands, where c) and d) correspond to the Wannier functions of the spin up bands (the spin down bands are localized on the other DBH of the unit cell with a similar distribution). The colours in panel b) stand for the spin sign, while in panels c) and d) colours stand for the sign of the orbital lobes.}
\label{fig3}
\end{figure}
     
In this section, we study the magnetic ground state of hypothetical DBH planar 2D crystals. We employed a plane waves spin-polarized DFT methodology with the PBE density functional and a scalar-relativistic PAW pseudopotential (see supp. mat. for details). Because we expect an antiparallel order, the unit cell consists of two DBH molecules, and the rotation operator ${\cal A}$ may acquire two possible values: $C_2$ or $C_4$. In all cases, we impose a planar geometry as a constraint, and coupling to adjacent DBH did not quench magnetism in the crystals. 

We first consider a DBH crystal with ${\cal A} = C_2$, where the DBH molecules are not rotated with respect to each other. The resulting centered rectangular lattice presents a different coupling in the horizontal and vertical directions, where the solution with the lowest energy hosted non-negligible local magnetic moments, zero net magnetization, a direct band gap, and no spin splitting in the Brillouin zone (Fig.\ref{fig2}a). The antiferromagnetic ground state was separated by $264$ meV from the ferromagnetic solution, and by $224$ meV from the non-magnetic phase. All of this indicates that this crystal is a collinear antiferromagnet with an intact Kramers degeneracy.  

Conveniently, it is also possible to perform a $\pi/2$ rotation in one of the DBH of the unit cell, producing a totally different platform that consists in a square lattice with ${\cal A}=C_4$ and an equivalent coupling in the four crystal directions (Fig.\ref{fig1}g). As before, this crystal presented a spin polarized lowest energy solution with compensated spin densities at adjacent DBHs, separated by $171$ meV ($346$ meV) from the ferromagnetic (non-magnetic) solutions. However, and different to the case with ${\cal A}=C_2$, the band structure showed a broken Kramers degeneracy for the selected $\bold{k}$-path (Fig.\ref{fig2}b), presenting an NRSS characteristic of an altermagnet. 
The spin-splitting in the bands close to the Fermi energy was $\sim25$ meV and $\sim 100$ meV, between one and two orders of magnitude smaller than those predicted for typical altermagnetic candidates with $d$-metals like \ch{RuO$_2$}\cite{vsmejkal2020crystal} or \ch{MnF$_2$}\cite{vsmejkal2022emerging}. 

Additionally, we represented constant energy contours for energies inside the first valence or conduction bands around $E_F$ (Fig.\ref{fig2}c,d). The broken ${\cal T}$, as a consequence of the $\pi/2$ rotated DBHs, causes a spin-splitting with a d-wave symmetry, similar to the rutile lattice at its Fermi energy\cite{vsmejkal2020crystal}. The results shown in this section indicate that AM can be obtained regardless the chemical nature of the spin centers, including $\pi$-magnetic carbon materials, thus expanding the list of candidates that are prone to host this new type of collinear magnetism. 

In Figure \ref{fig3}a, we show the tight-binding eigenvalues close to $E_F$ for the DBH graph, which can be calculated by diagonalizing the following spinless Hamiltonian with one $\pi$ orbital per atom:

\begin{equation}
{\cal H}_t = t \sum_{\langle i,j\rangle} (c^\dagger_i c_j + h.c.),
\end{equation}

where $c^\dagger_{i}$ creates and $c_i$ annihilates an electron at orbital $i$, the sum runs only through $\pi$ orbitals that belong to first-neighbouring atoms ($\langle i,j\rangle$) and hopping is accounted by the parameter $t$. At this level of theory, LUMO and HOMO orbitals ($\phi_{\pm}$) can be combined, obtaining back the wave functions of the two unpaired electrons from the open-shell configuration:

\begin{equation}
\varphi_{1,2}(i) = \frac{\phi_{+}(i)\pm\phi_{-}(i)}{\sqrt{2}}.
\end{equation}

In order to identify the role of these orbitals in the band structure, we computed the Wannier wave functions of the four highest occupied bands at half filling, showing a band dispersion in good agreement with the DFT calculation (Fig.\ref{fig3}b). As we can see in Figure \ref{fig3}c-d, the Wannier functions of the two spin up bands presented a considerable $\pi$ character, with an orbital distribution very similar to that of the $\varphi_z(i)$ unpaired electrons of isolated DBH molecules.    

\section{Covalent organic frameworks as altermagnetic candidates}

\begin{figure}
 \centering
    \includegraphics[width=0.5\textwidth]{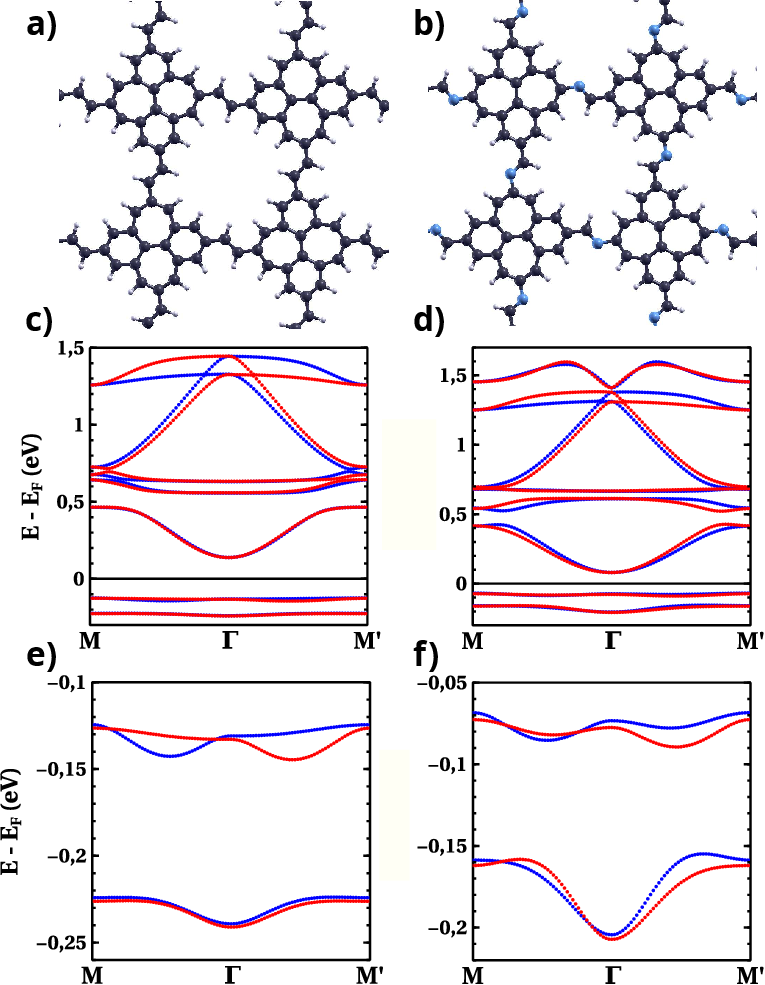}
\caption{Relaxed planar geometries of COFs made of DBH with a) -$\ch{C}$=$\ch{C}$- and b) -$\ch{C}$=$\ch{N}$- as linkers. c), d) DFT calculated band structures of crystals from panels a) and b), with $E_F = -3.02$eV and $E_F=-3.22$eV, respectively, and e), f) a zoom of the four highest occupied valence bands. Colours stand for the spin. Blue, dark, and white balls represent \ch{N}, \ch{C}, and \ch{H} atoms, respectively.}
\label{fig4}
\end{figure}

The formation of a potential altermagnetic 2D crystal of DBH molecules (Fig.\ref{fig1}g) might be rather elusive with conventional on-surface synthesis
experiments, as it might be difficult to control the relative orientation of adjacent monomers, hindering the altermagnetic phase. In the following, we explore the properties of related 2D crystals with suitably functionalized DBH to yield covalent organic frameworks (COFs) with DBH moieties. 

Following this, we have performed calculations using two different diatomic linkers, i.e., $\ch{C}{sp^2}$=$\ch{C}{sp^2}$ and $\ch{C}{sp^2}$=$\ch{N}{sp^2}$ (Fig.\ref{fig4}), testing the robustness of AM if additional symmetries are broken. 
We performed CASSCF-NEVPT2 calculations on the minimal dimer that conforms these COFs, consisting in two DBH molecules joined by a linker, obtaining the same order of many-body states as in Table \ref{table1}, but with reduced excitation energies because of the increased distance between the magnetic blocks (Table \ref{table2}). In the periodic crystals, according to DFT, an antiferromagnetic phase was also lower in energy than the non-magnetic and ferromagnetic solutions by $638$ meV ($526$ meV) and $15$ meV ($13$ meV), respectively, for the -$\ch{C}$=$\ch{C}$- (-$\ch{C}$=$\ch{N}$-) linkers, confirming an antiparallel magnetic ground state also in these systems.

\begin{table}[htp]
 \begin{center}
    \begin{tabular}{ c | c | c | c | c | c | c |}
    \cline{2-7}
   & \multicolumn{6}{c|}{$\Delta E$ (meV)} \\
    \cline{2-7}
   & \multicolumn{3}{c|}{BP86/6-31G**} & \multicolumn{3}{c|}{PBE0/def2-SVP} \\
    \hline
    \multicolumn{1}{|c|}{Linker} & $S=0$ & $S=1$ &  $S=2$ & $S=0$ & $S=1$ &  $S=2$ \\ \hline
    \multicolumn{1}{|c|}{-$\ch{C}$=$\ch{C}$-} & 0 & 3 & 7 & 0 & 3 & 7 \\ 
    \multicolumn{1}{|c|}{-$\ch{C}$=$\ch{N}$-} & 0 & 3 & 5 & 0 & 3 & 6 \\ 
    \hline
    
    \end{tabular}
    \label{table2}
\end{center}
\caption{Lowest energy many-body states referred to the ground state ($\Delta E$) for DBH dimers connected by two different linkers calculated with the CASSCF-NEVPT2 methodology. The geometries used were relaxed with DFT for the open-shell $S=0$ antiferromagnetic solution with ferromagnetic spin density at each DBH, using BP86/6-31G** and PBE0/def2-SVP (left and right columns, respectively).}\label{table2}
 \end{table}

The geometry of COFs from Figure \ref{fig4} is invariant under $C_4$, and adjacent DBH blocks can be related by a $\pi/2$ rotation, so altermagnetic features can be expected.
However, as we can see in Figure \ref{fig4}c-f, while keeping the NRSS from AM, there is an additional shift in the spin bands that breaks the spin degeneracy also at $\Gamma$. As it was explained in a recent work by L. Yuan et al.\cite{yuan2024nonrelativistic}, an NRSS can be induced at the center of the Brillouin zone if the spin centers cannot be related by any rotation symmetry, deviating from ideal AM where spin bands cross at that point.
In contrast, the designed COFs keep the $C_4$ rotation (${\cal A} = C_4$), but still display a spin splitting at $\Gamma$ (Fig.\ref{fig4}e-f), suggesting that the breaking of mirror symmetry by the linkers is enough to prevent the Kramers degeneracy at this point of the reciprocal space.
   
It is possible to demonstrate that a reduced symmetry may cause a shift in the, otherwise degenerate, spin up and down orbitals without the need of external magnetic fields or SOC. If we consider a different on-site energy on each sublattice\cite{castenetto2023edge}, then mirror symmetry is broken in a bipartite Hubbard Hamiltonian:

\begin{equation}
{\cal H} = {\cal H}_t + {\cal H}_U + \sum_{A,\sigma} \lambda_{A} c^\dagger_{A\sigma} c_{A\sigma}+ \sum_{B,\sigma} \lambda_{B} c^\dagger_{B\sigma} c_{B\sigma}, 
\label{eq4}
\end{equation}

where ${\cal H}_U=U\sum_i n_{i\sigma}n_{i\bar{\sigma}}$ is a Hubbard term, $\sigma\in \{\uparrow\downarrow\}$, and $\lambda_A\neq\lambda_B$. We also consider that these on-site energies are not sufficiently apart to quench magnetism and $U\gg |t|$, so the system is an antiferromagnet in the Mott regime in a mean-field approximation.
Such inequivalence in the on-site energy of adjacent $A$ and $B$ sites breaks mirror symmetry in any antiferromagnetic lattice with no frustration. 

In the simple case of a mean-field Hubbard dimer, the eigenvalues of Equation \ref{eq4} depend on the difference between both on-site energies:

\begin{equation}
\varepsilon_{\sigma\pm} = \frac{\lambda_A + \lambda_B + U\pm \chi_\sigma}{2},
\label{eq5}
\end{equation}

and

\begin{equation}
\chi_\sigma = \sqrt{\Delta_\lambda^2 + U(U+ 2\alpha\Delta_\lambda) + 4t^2},
\label{eq6}
\end{equation}

where $\Delta_\lambda = \lambda_A - \lambda_B$, and $\alpha$ is a sign that depends on $\sigma$. Hence, if the on-site energies are different ($\Delta_\lambda \neq 0$), and the electronic interactions are present ($U>0$), there is a spin splitting in the molecular orbital spectrum (see supp. mat.).

\section{Non-planar DBH systems}

\begin{figure}
 \centering
    \includegraphics[width=0.5\textwidth]{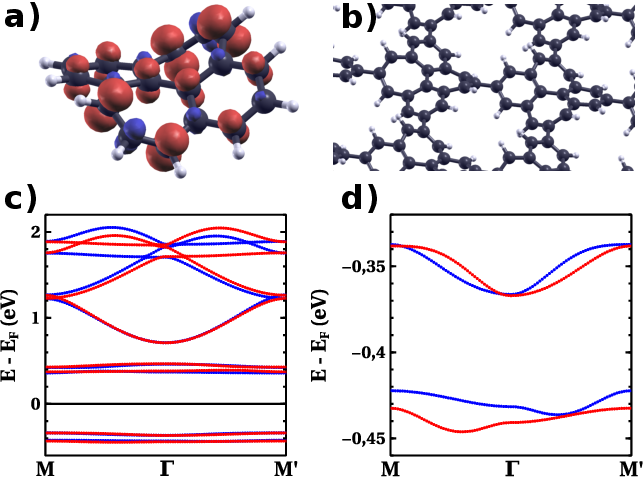}
\caption{a) Relaxed non-planar $S=1$ geometry, where red and blue stand for the sign of the spin density, done with DFT with the PBE0/def2-SVP density functional and basis. b) Relaxed geometry without constraints of the non-planar DBH crystal with compensated spin density, where c) and d) are its band structure and a zoom of the four highest occupied bands, done with periodic DFT with the PBE density functional and PAW pseudopotential for \ch{C} and \ch{H} atoms.}
\label{fig5}
\end{figure}

So far we have characterized the properties  of planar systems
based on DBH. However, the DBH planar geometry is a first-order saddle
point connecting equivalent conformations with opposite curvature. 
More specifically, according to DFT, a relaxed nonplanar
$S = 1$ geometry is more stable than the open-
/closed-shell $S = 0$ non-planar geometries ($\approx 40 - 200$
meV). CASSCF-NEVPT2 calculations confirm a
spin crossover where the many-body most stable wave function depends
strongly on the chosen geometry (Table \ref{table3}).

\begin{table}[htp]
 \begin{center}
    \begin{tabular}{ c | c | c | c | c |}
    \cline{2-5}
   & \multicolumn{4}{c|}{$\Delta E$ (meV)} \\
    \cline{2-5}
   & \multicolumn{2}{c|}{BP86/6-31G**} & \multicolumn{2}{c|}{PBE0/def2-SVP} \\
    \hline
    \multicolumn{1}{|c|}{Non-planar geometry} & $S=0$ & $S=1$ & $S=0$ & $S=1$  \\ \hline
    \multicolumn{1}{|c|}{Triplet} & 162  & 0  & 147 & 0  \\ 
    \multicolumn{1}{|c|}{Closed-shell} & 0 & 669  & 0 & 908  \\ 
    \multicolumn{1}{|c|}{Open-shell singlet} & 147 & 0 & 125 & 0  \\ 
    \hline
    
    \end{tabular}
    \label{table3}
\end{center}
\caption{Lowest energy many-body states referred to the ground state ($\Delta E$) for single DBH molecules with non-planar geometry calculated with the CASSCF-NEVPT2 methodology. The geometries used were relaxed with DFT for $S=1$ and open-/closed-shell $S=0$ magnetic configurations, using BP86/6-31G** and PBE0/def2-SVP (left and right columns, respectively).}\label{table3}
 \end{table}

\begin{figure}
 \centering
    \includegraphics[width=0.4\textwidth]{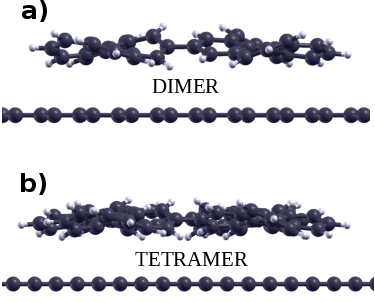}
\caption{Relaxed geometries of a) an open-shell $S=0$ DBH dimer and b) an open-shell $S=0$ tetramer supported both on graphene, done with the tight-binding 3OB method.}
\label{fig6}
\end{figure}
 
As a consequence, the ground state of DBH is a non-planar magnetic $S=1$ state with two parallel unpaired electrons (Fig.\ref{fig5}a). Interestingly, it is only structurally possible to link non-planar DBH molecules in a bidimensional periodic crystal if we use the $S=1$ geometry, while the closed-shell geometry does not permit to covalently bond tetrameric clusters. In Figure \ref{fig5}b, we show the relaxed geometry of a DBH crystal in a non-planar conformation, where the antiferromagnetic solution was lower in energy in a DFT calculation than the ferromagnetic and closed-shell solutions by $57$ meV and $544$ meV, respectively. The calculated band structure (Fig.\ref{fig5}c) presented an NRSS compatible with AM, but with a broken spin degeneracy at $\Gamma$ as in the case of COFs (Fig.\ref{fig5}d), in accordance with a lower symmetry originated by non-planarity.

Because conventional experiments on magnetic nanographenes are done supported on surfaces that tend to planarize the adsorbed molecules\cite{vegliante2024tuning,catarina2024conformational}, we performed geometry optimizations of DBH clusters adsorbed on graphene with DFT (see supp. mat.). We could confirm that finite DBH nanostructures (Fig.\ref{fig6}a,b) were not yet entirely planar but with a slightly lower dihedral angle of the bond that connects the different DBH in a tetramer ($\sim 45^\circ$) compared with the periodic system in vacuum ($\sim 50^\circ$). Magnetism is still present in these molecules, as it could be inferred from CASSCF-NEVPT2 calculations, where the monomer had an $S=1$ ground state separated by $156$ meV from the first excited $S=0$ state, indicating a modest stabilization of the triplet compared to the non-planar geometry relaxed in vacuum. Additionally, the dimer presented an open-shell $S=0$ ground state followed by an $S=1$ ($S=2$) excited states by $11$ meV ($33$ meV), similar to the planar dimers but with smaller excitation energies as a consequence of a weaker coupling due to the misalignment of the $\pi$ orbitals at the connecting sites. 

In order to finish, we wanted to mention that the non-planar $S=1$ DBH geometry presents planar chirality, thus this molecule is chiral if rotations are restricted to the plane. This feature opens the door to a richer geometrical behaviour of the non-planar clusters and crystals, where different diasteroisomers may differ in the strength of the magnetic coupling. In any case, a more exhaustive description, addressing the role of this feature in magnetic DBH nanostructures will be the aim of future work. 

\section{Conclusions}

We have studied the altermagnetic ground state of organic 2D crystals hosting DBH molecules as spin centers. Using CASSCF-NEVPT2 methodology, the ground state of planar DBH has been confirmed to be $S=1$, showing a robust antifferomagnetic coupling in DBHs dimers. 

The $D_{2h}$ symmetry of DBH permits the formation of a periodic square lattice with compensated local moments, where time reversal symmetry can be broken if one of the DBHs in the unit cell is rotated by $C_4$ with respect the other. The resulting crystal presented a DFT calculated band structure with an NRSS, characteristic of d-wave altermagnets. 
Keeping in mind that an experimental control of ${\cal A}\in \{C_2,C_4\}$ might be difficult, we calculated the magnetic properties of COFs consisting of DBHs linked by -$\ch{C}$=$\ch{C}$- or -$\ch{C}$=$\ch{N}$- moieties, also obtaining an NRSS, but with an additional broken spin degeneracy at the $\Gamma$ point because of their broken mirror symmetry. 

Finally, we noticed that the most stable geometry of DBH was a non-planar $S=1$, still allowing the formation of a 2D crystal with unquenched antiparallel magnetic moments. The calculated DFT band structure, in a similar way as the COFs, presented an NRSS with a non-degenerate $\Gamma$ point. These results pave the way for the realization of altermagnetism in organic bidimensional materials.    

\section{Acknowledgements}

This work was supported financially within the scope of the
project CICECO-Aveiro Institute of Materials, UIDB/50011/
2020 (DOI 10.54499/UIDB/50011/2020), UIDP/50011/
2020 (DOI 10.54499/UIDP/50011/2020), and LA/P/0006/
2020 (DOI 10.54499/LA/P/0006/2020), financed by national
funds through the FCT/MCTES (PIDDAC). This work has received funding from the
European Union’s Horizon 2020 research and innovation
program, under grant agreement No. 101046231,
and from the Foundation for Science and Technology (FCT)
under grant agreement M-ERA-NET3/0006/2021 through the
M-ERA.NET 2021 call.
K.S. acknowledges funding from the Scientific Employment Stimulus Program (2022.07534.CEECIND).

\bibliographystyle{apsrev-title}
\bibliography{references.bib}

\end{document}